\documentclass[graybox]{svmult}

\usepackage{mathptmx}       
\usepackage{helvet}         
\usepackage{courier}        
\usepackage{type1cm}        
\usepackage{makeidx}         
\usepackage{graphicx}        
\usepackage{multicol}        
\usepackage[bottom]{footmisc}

\makeindex             

\begin{document}

\title*{Avalanche dynamics and correlations in neural systems}

\author{Fabrizio Lombardi, Hans J. Herrmann and Lucilla de Arcangelis}
\institute{Fabrizio Lombardi \at Keck Laboratory for Network Physiology, Department of Physics, Boston University, Boston, Ma, USA \email{fabri@bu.edu}
\and Hans J. Herrmann \at Institute of Computational Physics for Engineering Materials, IfB, ETH Zurich, Switzerland \email{hans@ifb.baug.ethz.ch}
\and Lucilla de Arcangelis \at Department of Engineering, University of Campania ``Luigi Vanvitelli'', INFN sez. Napoli Gr. Coll. Salerno, 
Aversa(CE), Italy \email{lucilla.dearcangelis@unicampania.it}}
%
%
\maketitle

\abstract{The existence of power law distributions is only a first requirement in the validation of the critical behavior of a system.  Long-range spatio-temporal correlations are fundamental for the  spontaneous neuronal activity to be the expression of a system acting close to a critical point. This chapter focuses on temporal correlations and avalanche dynamics in the spontaneous activity of cortex slice cultures and in the resting fMRI BOLD signal. Long-range  correlations are investigated by means of the scaling of power spectra and of Detrended Fluctuations Analysis. The existence of $1/f$ decay in the power spectrum, as well as of power-law scaling in the root mean square fluctuation function for the  appropriate balance of excitation and inhibition suggests that  long-range temporal correlations are distinctive of "healthy brains". The corresponding temporal organization of neuronal avalanches can be dissected by analyzing the distribution of inter-event times between successive events. In rat cortex slice cultures this distribution exhibits a non-monotonic behavior, not usually found in other natural processes. Numerical simulations provide evidences  that this behavior is a consequence of the alternation between states of high and low activity, leading to a dynamic balance between excitation and inhibition that tunes the system at criticality. In this scenario, inter-times show a peculiar relation with avalanche sizes, resulting in a hierarchical structure of avalanche sequences.  Large avalanches correspond to low-frequency oscillations, and trigger cascades of smaller ones that are part of higher frequency rhythms. The self-regulated balance of excitation and inhibition observed in cultures is confirmed at larger scales, i.e. on fMRI data from resting brain activity, and appears to be closely related to  critical features of avalanche activity, which could  
play an important role in learning and other functional performance of neuronal systems.}


\section{Introduction}
\label{sec:1}

Critical systems are characterized by divergent correlations in space and time. 
As a consequence  such systems lack of a characteristic length scale, and the   statistics of events is governed by power laws. At the beginning of this century, a novel experimental approach has revealed that spontaneous cortical activity is organized in bursts made of neuronal avalanches \cite{plenz:pl03}. An avalanche is defined as an ensemble of neurons that fire close-in-time, namely at a temporal distance smaller than a given time interval, and is usually characterized by its size and duration. The particular definition of size and duration depends on the specific signal that is used to investigate spontaneous network activity. In general, those definitions provide a measure for the population of firing neurons and the time interval covered by their almost-synchronous firings. 

Neuronal avalanches have been first identified in the  organotypic cultures from coronal slices of rat somatosensory cortex \cite{plenz:pl03}, where they are stable for many hours \cite{plenz:pl04}. The size and duration of neuronal avalanches follow power law distributions with very robust exponents, which is a typical feature of a system acting in a critical state, where large fluctuations are present and the response does not have a characteristic size. The same critical behavior has been observed also {\it in vivo} on rat cortical layers during early post-natal development \cite{plenz:tbg}, and on  the cortex of awake adult rhesus monkeys \cite{pnas2}, using microelectrode array recordings, as well as on dissociated neurons from rat hippocampus and cortex \cite{maz,pas} or leech ganglia \cite{maz}. Recently, avalanche dynamics has been also identified in the resting state of the human brain by means of non-invasive techniques such as magneto-encephalography (MEG) \cite{shriki13}.

All these experimental results consistently exhibit power law size distributions decaying with the exponent -1.5, and  duration distributions that follow a power law with exponent -2. These exponents are consistent with the  universality class of the mean field branching process \cite{zap} and are therefore independent of the dimensionality of the system. This property, verified experimentally, is unusual in critical phenomena and it has been shown numerically to depend on the combined action of plastic adaptation and refractory time \cite{laurens}: Starting from a fully connected network, refractory time selects a preferential direction in the synaptic connection between two neurons, while the repeated adaptation of less used connections leads to the progressive depletion of loops in the network. As a result, the network becomes "tree-like" providing the mean field universality class.

Brain activity having features typical of systems at a critical point represents a crucial ingredient for learning. Indeed, a neuronal network model reproducing quantitatively the experimentally observed critical state of the brain,  is able to learn and remember logical rules including the exclusive OR, which has posed difficulties to several previous attempts \cite{pnas}. Learning occurs via plastic adaptation of synaptic strengths and exhibits universal features: The learning performance and the average time required to provide the right answer are controlled by the strength of plastic adaptation, in a way independent of the specific task assigned to the system. Interestingly, even complex rules can be learned provided that the plastic adaptation is sufficiently slow. The implemented learning dynamics is a cooperative mechanism where all neurons contribute to select the right answer. In fact, because the system acts in a critical state, the response to a given input can be highly flexible, adapting more easily to different rules. The analysis of the dependence of the performance of the system on the average connectivity confirms that learning is a truly collective process, where a high number of neurons may be involved and the system learns more efficiently if more branched paths are possible \cite{pnas}. The investigation of the learning performance in neuronal networks provides also insights into the role of inhibitory synapses. Performing more tasks in parallel is a typical feature of real brains characterized by the coexistence of excitatory and inhibitory synapses, whose percentage in mammals is measured to have a typical value of 20-30\%. 
By investigating parallel learning of more Boolean rules in neuronal networks \cite{opt}, it has been evidenced that that multi-task learning results from the alternation of learning and forgetting of the individual rules. Interestingly, a fraction of 30\% inhibitory synapses optimizes the overall performance,  since it guarantees, at the same time, the network excitability necessary to express the response and the variability required to confine the employment of resources.

Slow relaxation is a fundamental feature of systems acting at the critical point. Hence, temporal correlations are relevant over long time scales, and  give  rise to bursts of events. This phenomenology is distinctive of many natural phenomena, where not only power law distributions arise,  but also a complex temporal organization of events is observed. Examples are earthquakes and solar flares, which share a number of statistical laws evidencing the presence of temporal correlations between successive events \cite{physrep}.
Several statistical tools are currently  available for detecting the presence of such correlations and investigate their temporal range. In the case of brain activity, these methods can be applied either to the raw neuronal signal, defined as a continuous variable, or else to the sequence of neuronal avalanches, considered as a point-process in time. An intriguing question is indeed how to reconcile well-known properties of neuronal signals, such as oscillations with different  characteristic frequencies, with neuronal avalanches lacking characteristic  spatial and temporal scales \cite{pnas2,poil12,front14}. 

Such techniques, applied to numerical and experimental data from systems at different scales, constitute the necessary ingredient towards a definite assessment of criticality. In parallel, numerical models are of crucial importance to   identify the dynamic mechanisms controlling correlations, and to understand the role of criticality in brain functions.
In this chapter, we will present a summary of different statistical methods and their application for the investigation of avalanche dynamics in neural networks, with a particular focus on temporal correlations. Experimental results will be supported by numerical studies with the objective of providing a coherent understanding of the temporal features of neuronal activity.

\section{Avalanche activity and power spectra}
\label{sec:2}

A first indication of long-range temporal correlations is provided by the  observation of $1/f$ noise. Indeed, by the Wiener-Kintchine theorem \cite{pathria}, this implies that the time correlation function decays to zero in an infinitely long time interval. Analyses of experimental recordings of spontaneous neural activity, from EEG, to MEG and LFPs, generally  show a power law decay in the power spectral density (PSD), $S(f) \propto f^{-\beta}$, with superimposed peaks at the characteristic frequencies of dominant brain rhythms.
However, estimated values of the exponent $\beta$  vary over a rather broad interval,  and  appear to depend on brain areas and patient conditions. In particular, for healthy subjects $\beta$ takes values within the interval  $[0.8,1.5]$ \cite{novik97,deste06,deste10,pritchard92,zara97}, whereas it is much larger in epileptic patients \cite{he2014}.

To understand some of the basic mechanisms affecting the PSD scaling of neural signals, in this section   
we investigate the relationship between network inhibition, i.e. the percentage of inhibitory synapses, and the scaling exponent $\beta$ in a neuronal network of integrate and fire neurons driven by slow external stimulation that simulates  spontaneous activity. 
We consider neurons on a scale-free network and consistently show that inhibitory neurons determine the scaling behavior of the PSD: For a neuronal network of only excitatory neurons, the PSD follows a power law with an exponent $\beta \simeq 2$. By introducing inhibitory neurons, $\beta$ decreases, and exhibits values in the interval $[1,1.4]$ for a percentage  of inhibitory synapses between 20 and 30\%, in agreement with experimental findings \cite{novik97,deste06,deste10,pritchard92}.


\subsection{Neuronal model and avalanche activity}
\label{model}

We simulate the intrinsic activity of a neuronal network by means of an integrate and fire model \cite{prl1,pre,chaos} inspired in self-organized criticality (SOC) \cite{jsn:soc}. In this model, each neuron $i$ is characterized by a membrane potential $v_i$ and fires if and only if $v_i$ is equal or above a firing threshold $v_{c}$. To trigger activity, namely to bring a neuron at or above the firing threshold, we apply a small stimulation to a randomly chosen neuron. Then, whenever at time $t$ the potential of neuron $i$ fulfils the condition $v_i \geq v_{c}$, the neuron fires, changing the potential of all connected neurons.\\
The $N$ neurons are located at the nodes of a network which can have a very general topology. In the following, we present results obtained for neurons randomly distributed in a square and  connected by a scale-free network of directed synapses. More precisely, to each neuron $i$ we assign an out-going connectivity degree, $k_{out_i}\in [2,100]$, according to the degree distribution $P(k_{out})\propto k_{out}^{-2}$ of the functional network  measured in  \cite{chia:sfn}, and two neurons are then connected with a distance-dependent probability, $P(r) \propto e^{-r/r_0}$, where $r$ is their Euclidean distance \cite{roerig02} and $r_0$ a typical edge length. To each synaptic connection we assign an initial random strength $g_{ij} \in [0.15, 0.3]$ and to each neuron an excitatory or inhibitory character by fixing a percentage $p_{in}$ of inhibitory synapses. Outgoing synapses are excitatory if their presynaptic neuron is excitatory, inhibitory otherwise. Since synapses are directed, $g_{ij} \neq g_{ji}$, in general out-degree and in-degree of a neuron do  not coincide. Therefore once the network of out-connections is established, we identify the resulting degree of in-connections, $k_{in_j}$, for each neuron $j$, namely we identify the number of synapses directed to $j$. 

The change in the  membrane potential of the postsynaptic neuron $j$ due to the firing of neuron $i$ is proportional to the relative synaptic strength $g_{ji}/\sum_l g_{li}$
 
\begin{equation}
v_j(t+1)=\ v_j(t) \pm \frac{v_i\cdot k_{out_i}}{k_{in_j}} \frac{g_{ji}}{\sum_{l=1} ^{k_{out_i}} g_{li}}.
\label{eqn:dv}
\end{equation}
\noindent
Here the normalization by the synaptic strengths ensures that during the propagation of very large avalanches the membrane potential assumes finite values, while the plus or minus sign is for excitatory or inhibitory synapses, respectively. After firing, the membrane potential of the neuron is set to $v_i = 0$ and the neuron remains in a refractory state for $t_{ref} = 1$ time step, during which it is unable to receive or transmit any signal. Each neuron in the network is an integrate and fire unit, therefore it will change its potential by summing the successive stimulations from presynaptic firing neurons according  to Eq. \ref{eqn:dv}. With respect to traditional neuronal models implementing  partial differential equations for the temporal dependence of the membrane potential, this model is a cellular automaton where time is a discrete variable. The time unit is the time delay between the triggering of the action potential in the presynaptic neuron and the change of the membrane potential in the postsynaptic neuron, which corresponds to few ms in real neuronal system. Within this temporal window our model is unable to provide the state of the neuron, however it is numerically very efficient thus allowing simulations of very large systems. 

When a neuron $i$ fires, its out-going connections induce a potential variation in the $k_{out_i}$ postsynaptic neurons. The strength $g_{ji}$ of these active synapses is increased  proportionally to the membrane potential variation $|\delta v_j|$ that occurred at the postsynaptic neuron $j$

\begin{equation}
g_{ji}=\  g_{ji} + | \delta v_j |/v_{c}.
\label{eqn:spla}
\end{equation}                                    

Conversely, the strength of all inactive synapses during an avalanche is reduced by the average strength increase per bond

\begin{equation}
\Delta g=\sum_{ij} \delta g_{ji}/N_B,  
\label{eqn:dg}
\end{equation}
\noindent
where $N_B$ is the number of bonds. We set a minimum and a maximum value for the synaptic strength $g_{ij}$, $g_{min}=0.0001$ and $g_{max}=1.0$. Whenever $g_{ij} < g_{min}$, the synapse $g_{ij}$ is pruned, i.e. permanently removed.

\begin{figure}[h!]
\begin{center}
\includegraphics[width=8.0cm]{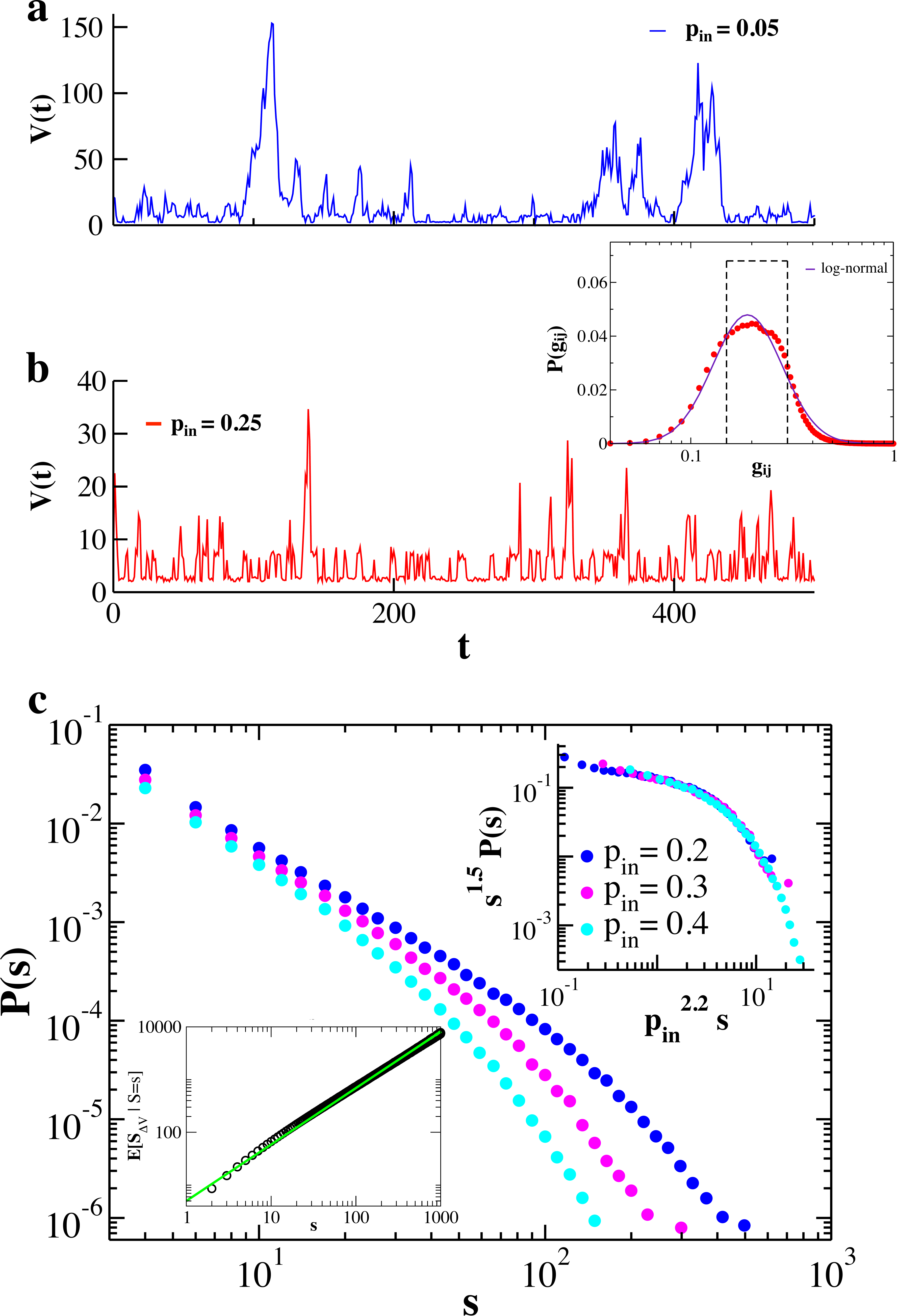}
\end{center}
\caption{Network activity and avalanche size distributions  for different fraction $p_{in}$ of inhibitory synapses. The intensity of activity $V(t)$ is the sum of all potential variations occurring in the network at each time step, namely $V(t) = \sum_i \delta v_i (t)$. Network activity for $p_{in}=0.05$ a) and $p_{in}=0.25$ b). Inset of b): Distribution $P(g_{ij})$ of synaptic strengths at the end of the initial period of plastic adaptation. The dashed line indicate the initial uniform distribution; c) Avalanche size distributions on a scale-free network for different values of $p_{in}$. Upper inset:  Plotting $ s^{\alpha} P(s)$ vs $p_{in} ^{\theta} s$, with $\alpha = 1.5$ and $\theta = 2.2$, data collapse onto a universal scaling function; Lower inset: Relation between  the number of firing neurons $s$ and the sum $s_{\Delta V}$ of all positive potential variations (depolarizations)  $\delta v_i ^{+}$ occurred in the network, namely  $s_{\Delta V} = \sum_i \delta v_i ^{+}$.}
\label{fig:signal}
\end{figure}

These rules constitute a Hebbian-like scheme for synaptic plastic adaptation. The network memorizes the most used synaptic paths by increasing their strengths, whereas less used synapses eventually atrophy. They implement a sort of long-term synaptic plasticity, a homeostatic mechanism that dynamically balances synaptic strengthening and weakening in the network. Short-term plasticity is not taken into account in the present version of the model.
Plastic adaptation is implemented during an initial sequence of external stimuli and shapes the network of synaptic strengths. Its extent can be viewed as a measure for the experience of the trained network. Synaptic strengths are initially uniformly distributed in the interval $[0.15,0.3]$. The distribution of $g_{ij}$ resulting from plastic adaptation is shown in the inset of Fig. \ref{fig:signal}b, and closely resembles the distribution of synaptic strengths measured experimentally \cite{gg_lognorm}.
Then avalanche activity is measured  for fixed synaptic connections by applying a new sequence of random stimulations. Figures \ref{fig:signal}a and b show the network activity as function of time for different percentage of inhibitory neurons.

An avalanche is defined as a cascade of successively firing neurons after an external stimulation and  can involve a variable number of neurons. The avalanche size  is defined as the number of firing neurons $s$, or, alternatively, as the sum $s_{\Delta V}$ of all positive potential variations (depolarizations)  $\delta v_i ^{+}$ that occurred in the network, namely  $s_{\Delta V} = \sum_i \delta v_i ^{+}$. Avalanches are also characterized by their duration $T$, which is defined as the number of iterations taken by the activity propagation. 
The distributions of avalanche sizes, $P(s)$, and durations, $P(T)$, obtained by this model, are in agreement with experimental data \cite{plenz:pl03,shriki13}, namely they exhibit a power law behavior with exponents $\alpha \simeq -1.5$ and $\tau\simeq -2.0$, respectively, followed by an exponential cutoff (Fig. \ref{fig:signal}c and Fig. \ref{spectra}b). As shown in the lower inset of Fig. \ref{fig:signal}c, a scaling relation exists between $s$ and $s_{\Delta V}$, namely $s_{\Delta V}(s) \sim s^{\gamma_s}$ with $\gamma _s \simeq 1$, implying that $P(s_{\Delta V})$ follows the same scaling behavior as the distribution of avalanche sizes measured in terms of firing neurons. 

The scaling behavior of such distributions is independent of the network topology  and of the percentage $p_{in}$ of inhibitory synapses. On the other hand, $p_{in}$ significantly affects the exponential cutoff, as shown in Fig. \ref{fig:signal}c: 
By increasing the percentage of inhibitory synapses  the exponential cutoff gradually moves towards smaller avalanche sizes $s$ and the scaling  regime  shrinks, evidencing the universal scaling behavior $P(s) \sim s^{-\alpha} f(s/p_{in} ^{-\theta})$. The scaling function confirms that the size of the largest possible avalanche decreases with the percentage of inhibitory synapses, whereas the critical exponent $\alpha$ shows a stable value 1.5. Similar behavior is observed for the duration distributions \cite{chaos}.

\subsection{Power spectra}
\label{spectra}

The analysis of the size and duration distributions evidences the crucial role played by inhibition in avalanche dynamics \cite{chaos}. A very first step in the investigation of temporal correlations in neuronal signals is the analysis of the  power spectrum \cite{chaos}. In real neuronal network, one generally studies the electric signal resulting from the firing of a more or less localized population of neurons, such as local field potentials (LFPs) \cite{plenz:pl03}. In a  close analogy with LFPs, for each network configuration we construct a temporal signal,  $V(t)$, as the sum  of all potential variations occurring at each time step of network activity, i.e  $V(t) = \sum_i \delta v_i (t)$. We then evaluate the Fourier transform of this signal, whose squared amplitude averaged over all configurations provides the PSD of the network activity. 
The signal $V(t)$ significantly depends on $p_{in}$, and exhibits not only smaller peaks,  but also sparser high amplitude fluctuations for larger percentage of inhibitory synapses, as shown in Fig. \ref{fig:signal}a. 
The dissipative property of inhibitory synapses turns out to have a strong influence on  the PSD of avalanche activity, $S(f)$. In Fig. \ref{spectra}a, we  show  $S(f)$ for several values of $p_{in}$.  
\begin{figure}
\begin{center}
\includegraphics[width=11.0cm]{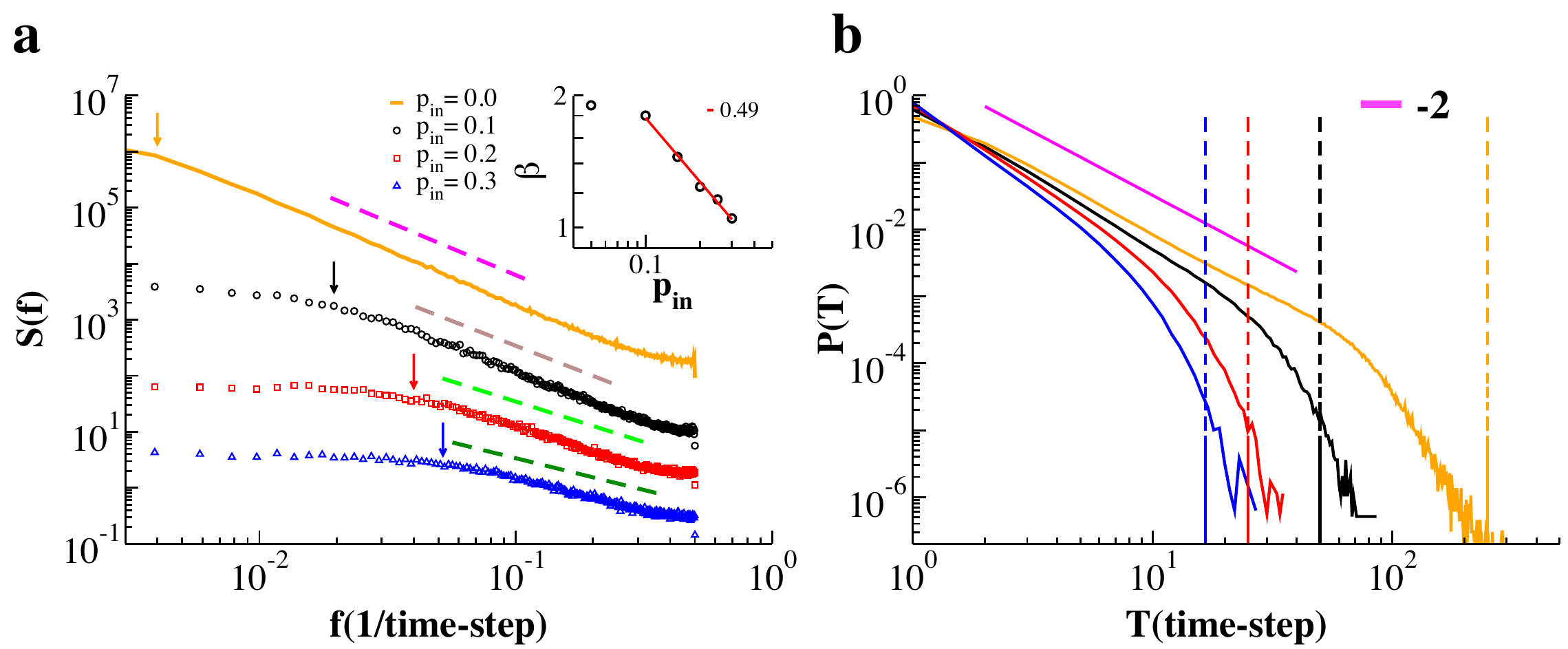}
\end{center}
\caption{Power spectral density of avalanche activity (a) and duration distribution $P(T)$ (b) for a scale-free network with $N=64000$ neurons and different $p_{in}$. The PSD follows a power law whose exponent $\beta$ depends on $p_{in}$ (Inset) and approaches the value $\beta = 1$  for $p_{in}=0.3$. The cutoff at low frequencies (arrows), which indicates the transition to white noise, scales with $p_{in}$ and corresponds to avalanche durations in the exponential cutoff of $P(T)$ (dashed lines). }   
\label{spectra}
\end{figure}

In the simulations the time unit can be roughly estimated as the time interval  between the firing of the presynaptic neuron and the induced voltage change in the postsynaptic neuron,  which in real systems should correspond to a few milliseconds. With this rough correspondence, our PSD frequency range goes from approximately 1 to 100 Hz.
We notice that the PSD has the same qualitative behavior for each value of $p_{in}$, namely a $f^{-\beta}$ decay and a cutoff at low frequencies that, as the cutoff in the avalanche size distribution shifts towards smaller $s$ values, moves towards higher frequencies with increasing $p_{in}$. The low-frequency cutoff indicates the transition to white noise, which characterizes an uncorrelated process, and  corresponds to avalanche durations in the exponential cutoff of the distribution $P(T)$ of avalanche durations. 
The exponent $\beta$ decreases for increasing values of the fraction of inhibitory synapses. In particular, for a purely excitatory network we find that $\beta$ is  close to 2, an exponent associated to the PSD of brown noise and larger than values found in experimental studies of spontaneous cortical activity in healthy subjects \cite{novik97,deste06,deste10,pritchard92}, but close to values measured for epileptic patients \cite{he2014}. 
However, when $p_{in}$ becomes closer to the fraction of inhibitory  synapses characteristic of real neuronal systems, i.e. about $0.3$, we find that $\beta$ is in the interval $[1,1.4]$, the range of  experimentally measured $\beta$ values \cite{novik97,deste06,deste10,pritchard92}. 
More specifically, for $0.1 \le p_{in} \le 0.3$ the exponent $\beta$ decays as $p_{in} ^{-\delta}$, where $\delta \simeq 0.49$, and tends to one as $p_{in}\to 0.3$ (Fig. \ref{spectra}a).
The range of the scaling regime varies as a function of $p_{in}$, and, according to our rough estimate of numerical time units, goes from few to nearly 100 Hz for  $0.1 \le p_{in} \le 0.3$ (Fig. \ref{spectra}a). The low-frequency cutoff (arrows in Fig. \ref{spectra}a) is indeed at about 4Hz, 8Hz, and 10Hz, for $p_{in} = 0.1, 0.2,$ and $p_{in} = 0.3$, respectively. As shown in \cite{chaos}, for a fixed value of $p_{in}$ the low-frequency cutoff scales with the systems size, and decreases for increasing sizes. 

The $1/f$ scaling regime observed in the resting state brain activity approximately  spanned the interval $[1,10]$Hz in \cite{pritchard92}, while more recent studies evidenced that this regime extends to higher frequencies \cite{novik97,deste10}, and shows superimposed peaks corresponding to dominant  brain rhythms \cite{buz2004}. In particular, \textit{Novikov} et al showed that the $1/f$ behavior extends up to 50Hz in the MEG of resting human brain, a range that considerably overlap with our numerical results. 
Moreover, it is interesting to notice that experiments on dissociated networks of rat hippocampal neurons and leech ganglia \cite{maz} exhibit a very similar behavior as a function of network inhibition: The power spectra for both systems exhibit a $1/f$ power law decay, however if inhibition is hindered by introducing  picrotoxin (PTX) or bicuculline in the physiological bath, the power-law decay becomes closer to brown noise, i.e. $1/f^{-2}$, as obtained in simulations of purely excitatory networks.

The scaling exponent $\beta$ of the PSD is related to the critical exponent $\frac{\tau -1}{\alpha -1}$, which  connects avalanche sizes and durations \cite{jsn:soc}, $s(T)\sim T^{\frac{\tau -1}{\alpha -1}}$. It has been shown that, for purely excitatory models with $\alpha < 2$, $\beta = \frac{\tau -1}{\alpha -1}$ \cite{sethna00}. Our model very closely follows this analytical prediction: Indeed for a purely excitatory network we find $\beta \simeq 2$ and $\frac{\tau -1}{\alpha -1} \simeq 2$. On the other hand, to our knowledge no analytical derivation of the relation between $\beta$ and $\frac{\tau -1}{\alpha -1}$ is available for systems with excitatory and inhibitory interactions. Deriving such a relationship for SOC-like models with inhibitory interactions is a general problem of great interest and implies the introduction of anti-ferromagnetic interactions in the model in  \cite{sethna00}.
The $1/f$ behavior of power spectra provides a first important evidence for the existence of long-range correlations in systems that are self-tuned into a critical state, and appears to be connected to the right percentage of inhibition in the system. In particular, \textit{de los Rios} et al. \cite{rios99} have shown that a dissipative term in the dynamics of the original sand pile model gives rise to avalanche activity whose PSD decays as $1/f$.  However, in their model the avalanche sizes are not distributed according to a power law. In our neuronal model instead, inhibition gives rise to $1/f$ power spectra and  avalanche size  distributions still follow a power law behavior with a universal scaling function.

\section{Inter-event time distributions}
\label{sec:3}
For a wide variety of natural stochastic phenomena the presence and extent of temporal correlations is analyzed in terms of the distribution of inter-event times. This is the distribution of time interval durations between successive events.
The advantage of this analysis is that the distribution of inter-event times has a simple exponential decay for a pure Poisson process, whereas it exhibits a power law regime over the temporal range where  correlations are relevant. In particular, it has been shown analytically that the Omori temporal decay of the earthquake rate after the occurrence of a large mainshock gives rise to a power law decay in the inter-event time distribution with an exponent related to the Omori exponent \cite{Utsu}. More generally for several processes, such as earthquakes, solar flares,or  acoustic emissions due to rock fracture, this distribution is monotonic and behaves as a Gamma function, exhibiting an initial power law decay followed by an exponential cutoff. This behavior is robust and universal with respect to different event catalogs and the lower threshold imposed to the event size in the temporal sequence \cite{Corral,physrep}.
Following this approach, we identified neuronal avalanches in rat cortex slice cultures and measured the distribution $P(\Delta t)$ of time intervals separating consecutive avalanches \cite{prl12}. Interestingly,  the inter-avalanche time  distribution exhibits a novel, non-monotonic behavior in different culture samples, with common features:
An initial power-law regime that is  characterized by exponent values between -2 and -2.3, a local minimum located at 200 ms $< \Delta t_{min} <$ 1 s, and a more or less pronounced maximum at $\Delta t \simeq 1-2$ s. This complex behavior is not usually observed in other natural phenomena and suggests that temporal correlations are not only relevant in avalanche occurrence, but reflect complex underlying dynamical mechanisms.

\subsection{Up-states and down-states}

Numerical simulations evidence that a traditional integrate and fire neuronal model is unable to reproduce the non-monotonic behavior of the inter-avalanche event distributions found experimentally. To achieve this goal we hypothesize that non-monotonicity arise from the transition between two substantially different network states, and introduce in the  model (\ref{model}) the concept of up and down-states.

Spontaneous activity exhibits a complex alternation between bursty periods, called up-states, where neuronal avalanches are detected, and quiet periods, named down-states. These are characterized by a general disfacilitation in the system, i.e. absence of synaptic activity, causing long-lasting returns to resting potentials in a large population of neurons \cite{wilson}. Action potentials are rare during down-states, however small amplitude depolarizing potentials originating from spontaneous synaptic release, may occur. The non-linear amplification of small amplitude signals leads to the generation of larger depolarizing events, bringing the system back into the up-state. Moreover, together with this network features, experimental observations indicate that neurons are characterized by two preferred values of the membrane potential: A very negative one, below resting potential, in the down-state, and a more depolarized one in the up-state. Since the neuron up-state is just a few millivolts below the action potential threshold, during the up-state neurons respond faster to synaptic stimulations, giving rise to close-in.time avalanches. The down-state instead is characterized by long periods of quiescence during which the network recovers from biomolecular mechanisms that hamper activity, as the exhaustion of available synaptic vesicles, the increase of the nucleoside adenosine inhibiting glutamate   release, or the blockade of receptor channels by the presence of external magnesium \cite{wilson,tim:00,tim:01,thom:aden}.

We implement up and down-states in the original model by monitoring the avalanche activity \cite{prl12}. We measure the size of each avalanche in terms of  depolarizations $\delta v_i$ of all active neurons, $s_{\Delta v}$ (see Sec. \ref{model}). As soon as an avalanche is larger than a threshold value, $s_{\Delta v}^{min}$, the system transitions into a down-state and neurons active in the last avalanche become hyperpolarized proportionally to their previous activity, namely we reset

\begin{equation}
v_i = v_i - h \delta v_i
\label{eqn:down}
\end{equation}
\noindent 
where $h > 0$. This equation implies that each neuron is hyperpolarized proportionally to its previous activity, i.e. its potential is the lower, the higher its potential variation in the previous avalanche. This rule introduces a short-range memory at the level of a single neuron and models a number of possible mechanisms for local inhibition \cite{tim:00,tim:01,thom:aden}.
Conversely, if an avalanche has a size smaller than $s_{\Delta v}^{min}$, the system remains or transitions into an up-state. In the up-state, the potentials $v$ of all neurons firing in an avalanche are not set equal to zero resting potential, but to the depolarized value 

\begin{equation}
v_i=v_c(1-s_{\Delta v}/s_{\Delta v}^{min})
\label{eqn:up}
\end{equation}

\begin{figure}[t!]
\begin{center}
\includegraphics[width=10.5cm]{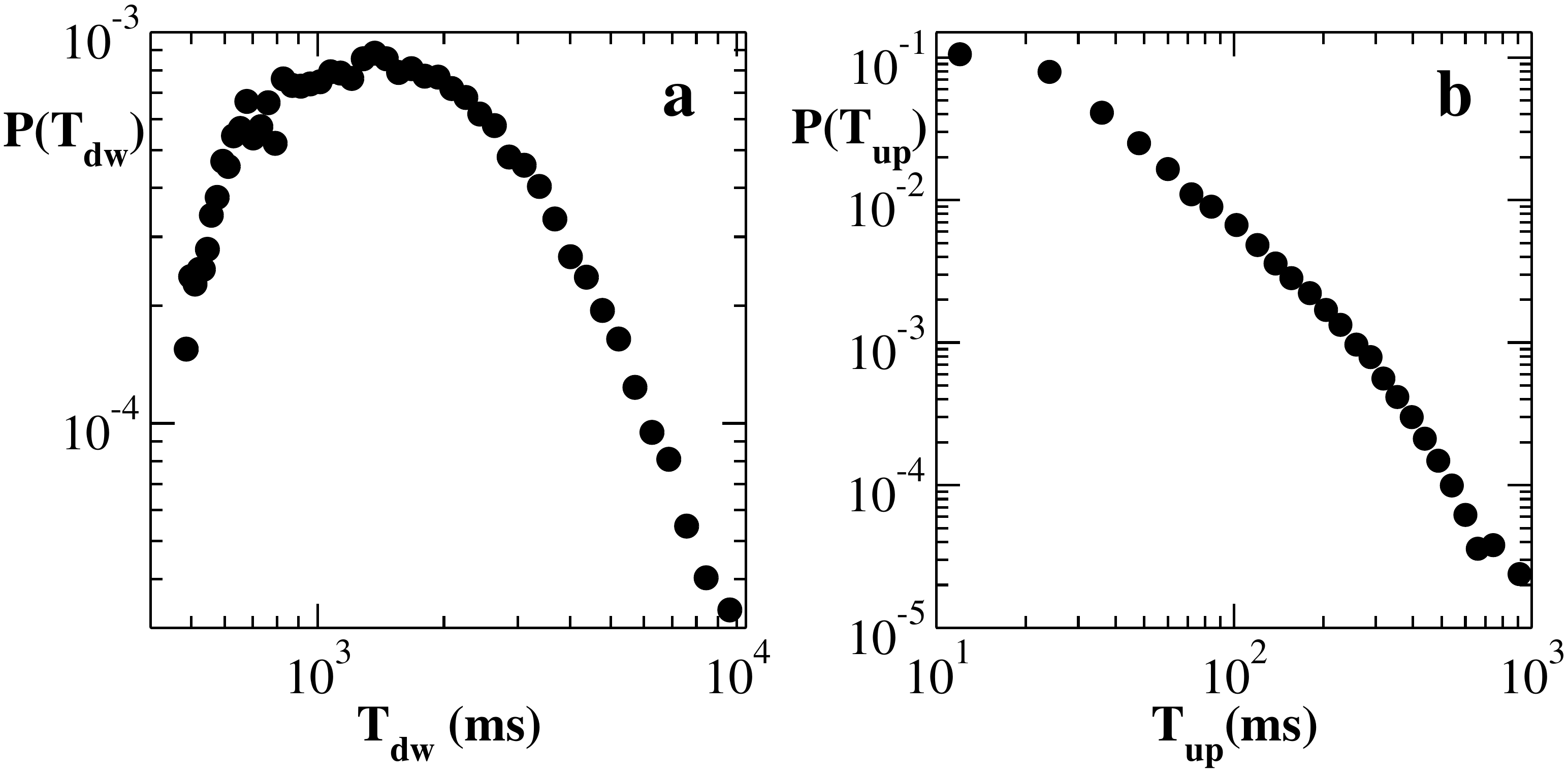}
\end{center}
\caption{Distribution of durations of down-states (a) and up-states (b) averaged over 100 configurations of networks with $N = 64000$ neurons with $p_{in} = 0.1$. }
\label{updw}
\end{figure}

The neuron membrane  potential in the up-state then depends on the response of the whole network and remains close to the firing threshold. The threshold $s_{\Delta v}^{min}$ controls the extension of the up-state and therefore the level of excitability of the system, whereas $h$ controls the level of hyperpolarization of neurons in the down-state and therefore its duration. Interestingly, the last two equations each depend on a single parameter, $h$ and $s_{\Delta v}^{min}$, which introduce a memory effect at the level of single neuron activity and of the entire system, respectively. These two parameters can be tuned separately in numerical simulations to obtain the best fit with experimental data.
The implementation of up and down-states in numerical simulations provides a temporal activity made of bursts of avalanches (up-states) followed by long periods of quiescence (down states). In Fig. \ref{updw} the duration distributions of up and down-states indeed show that down-states can last much longer that up-states, in agreement with experimental data on rat visual cortex \cite{coss03} and simulations of integrate and fire neuron networks \cite{mill}. 

\begin{figure}[t!]
\begin{center}
\includegraphics[width=11.5cm]{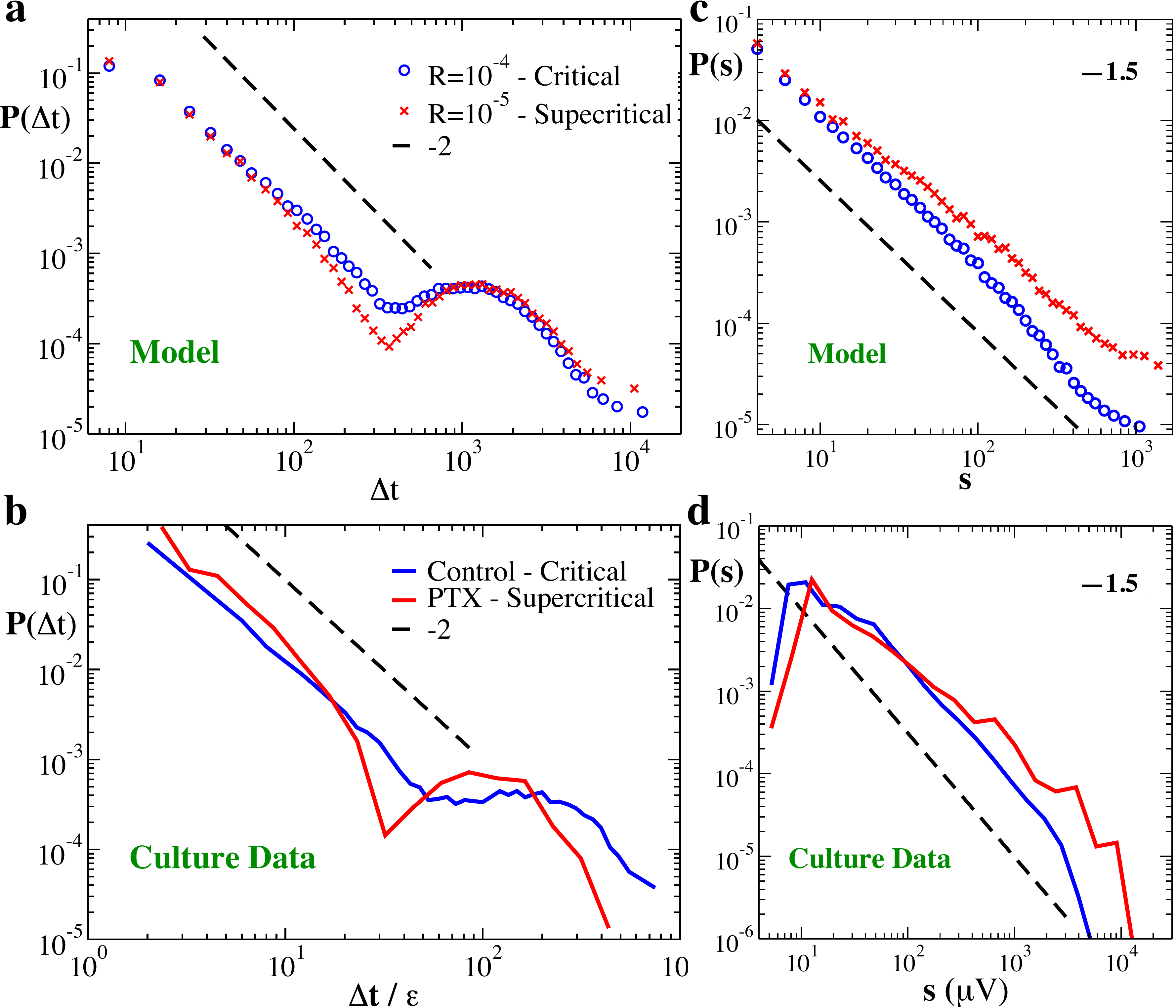}
\end{center}
\caption{Numerical and experimental distribution of inter-event time and avalanche size in critical and supercritical state. (a) Numerical inter-event time distributions averaged over 100 scale-free network configurations with $N = 16000$ neurons for values of $s_{\Delta v}^{min}$ and $h$ providing the best agreement  with experimental data in normal (critical) condition with $R \simeq 10^{-4}$ (blue circles). For smaller values of $R$ activity becomes supercritical (red crosses) and the distribution becomes more similar to data from disinhibited cultures. (b) Experimental inter-event time distribution for a slice of rat cortex in normal condition (blue line) and disinhibited by PTX (red line). (c) Numerical avalanche size distribution  in a  critical condition (blue circles), $R = 10^{-4}$, and in a super-critical condition, $R = 10^{-5}$ (red squares). (d) Experimental avalanche size distribution measured for the culture in normal, critical condition (blue line) and for the culture treated by PTX, i.e. in disinhibited,  super-critical condition (red line).}
\label{fig:ndt}
\end{figure}

The inter-avalanche time distribution for the temporal sequence of up and down-states is shown in Fig. \ref{fig:ndt}, together with experimental results  for critical and super-critical samples of rat cortex slices.  Super-critical samples are generally obtained by suppressing inhibition, or enhancing excitation. 
They are characterized by an excess of large avalanches and  a smaller $\alpha$ power law exponent in the avalanche size distribution as compared to the normal, critical samples (Fig. \ref{fig:ndt}d). 
The non-monotonic behavior of experimental data can be reproduced with very good agreement by tuning separately the two parameters $h$ and $s_{\Delta v}^{min}$. 
The close-in time avalanches occurring in the up-state are responsible for the initial power law regime with an exponent close to -2 (Fig. \ref{fig:ndt}a), as for experimental data (Fig. \ref{fig:ndt}b), evidencing that temporal correlations are extended over a range at least comparable to the up-state duration.
Conversely, the long-lasting down-states originate the bell-shaped behavior at large $\Delta t$ centered at a value that depends on $h$ and $s_{\Delta v}^{min}$. 
The superposition of these two different regimes then provides the local minimum and therefore the non monotonic behavior of the distribution.
  
The overall behavior of the inter-event time distribution, and thus the  temporal organization of avalanches, is controlled by a single parameter, the ratio  $R = h/ s_{\Delta v}^{min}$, which expresses the balance between excitation and inhibition, dynamically realized by the alternation of up and down-states.
By increasing the level of excitability, the system becomes super-critical, and    the distribution of avalanche sizes closely resembles the one for disinhibited cultures (Fig. \ref{fig:ndt}c). At the same time, we observe major changes in the inter-event time distribution (Fig. \ref{fig:ndt}a). In particular, the power law regime becomes shorter, its exponent changes, and the local minimum gets more pronounced. Similar changes are observed in cultures treated with PTX (Fig. \ref{fig:ndt}b) \cite{scirep16}. 

The model described here, consistently connects avalanche statistics with dynamical properties of neural systems that are captured by the inter-event time distribution. Very similar values of $R$ have to be implemented to reproduce the experimental distributions for different critical samples. This means that the up and down-state features consistently change to produce a specific  temporal organization of avalanches at criticality. In the next Section we will discuss the intimate connection between inter-event time and avalanche size.

\subsection{Avalanches and oscillations}

\begin{figure}
\begin{center}
\includegraphics[width=11.5cm]{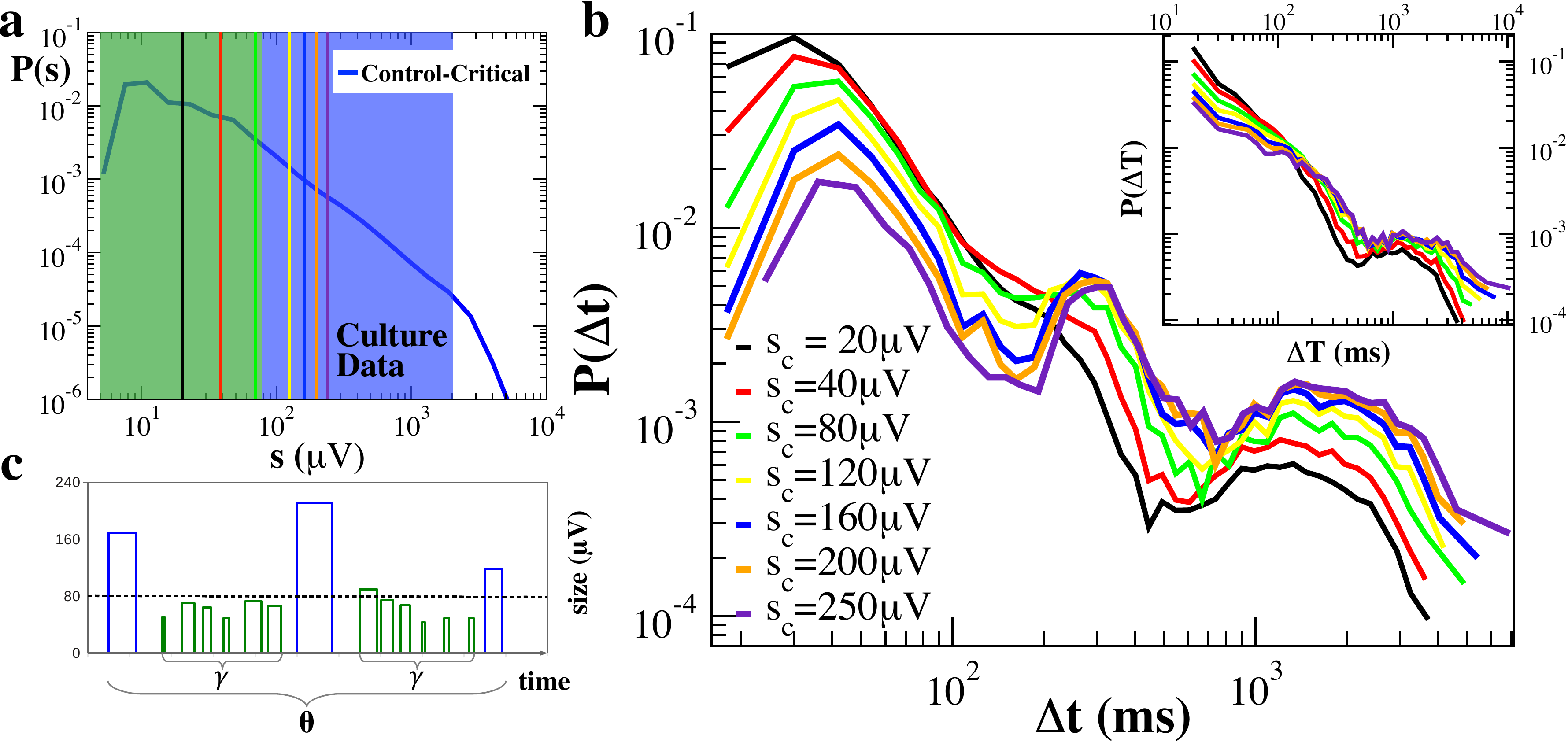}
\end{center}
\caption{(a) Experimental inter-event time distribution $P(\Delta t; s_c)$ are  evaluated by setting different  thresholds $s_c$ for the minimum avalanche size. The color code of the vertical bars corresponds to the legend of panel (b). Avalanches whose size falls in the blue region tend to occur at frequency $\theta$ or smaller, whereas avalanches with size in the green region tend to occur at higher frequency. (b) Experimental inter-event time distributions for different values of the threshold $s_c$ on avalanche size. Already for $s_c = 80 \mu V$, the distribution clearly exhibits an additional peak. Beside the one at large time scales, $\Delta t \simeq 1000 - 2000$ ms, which is related to the characteristic time of up-state recurrence, the peak at about 300 ms corresponds to the period of  $\theta$ oscillations, and the peak around 40 ms to the $\beta/\gamma$ rhythms. Inset: The same distributions evaluated for the reshuffled avalanche time series and for different values of the threshold $s_c$ on avalanche size. (c)
Hierarchical organization of avalanches with different sizes (bar heights)
 corresponding to temporal organization of nested $\theta -\beta /\gamma$ oscillations. Large avalanches (blue bars) occur with $\theta$ frequency and trigger smaller avalanches related to faster $\gamma$ oscillations (green bars).  Here bar widths indicate durations.  Spacing between blue bars corresponds to a $\theta$ period. Spacing between the starting points of green bars corresponds to the $\gamma$ period. Sizes $s$ of avalanches related to $\theta$ cycles tend to fall within the blue region of the size distribution $P(s)$ plotted in (a), whereas the sizes corresponding to nested $\gamma$ oscillations fall within the green region.}   
\label{ndt_th}
\end{figure}

According to similar studies performed for other dynamical processes, we analyze the experimental inter-event time distribution by constructing temporal sequences of events containing only avalanches whose size is larger than a given threshold $s_{c}$. Usually this selection provides a sparser sequence, where avalanches are larger and  more distant in time. In many cases, the scaling behavior of the inter-event time  distributions obtained using such a procedure for different thresholds turns out to be universal if the inter-event time  $\Delta t$ is rescaled by the average rate $\tau_0$, namely all distribution collapse onto a universal function $f$ \cite{Corral}, 
\begin{equation}
P(\Delta t; s_{c})=\tau_0 f(\tau_0\Delta t)
\label{eqn:corral}
\end{equation}
where $f$ is a Gamma function. This result evidences the presence of a unique time scale in the process, namely the average inter-event time or the inverse average rate.
The same analysis performed on experimental data for seven samples of rat cortical slices \cite{front14} evidences a more complex behavior, as shown in Fig. \ref{ndt_th} for the critical sample discussed in Sec. \ref{sec:3}. 
By increasing the minimum size threshold from 20 $\mu$V to 250 $\mu$V, the distributions, as expected, show a decrease in the probability to observe small $\Delta t$ and a corresponding increase at large $\Delta t$. More strikingly, the functional form of the probability distribution is clearly non-universal, since for increasing $s_c$ values pronounced peaks emerge at $\Delta t$s corresponding to the characteristic  periods of  $\theta$ and $\beta/\gamma$ rhythms, $\Delta  t_{\theta} \simeq 300$ms and $\Delta  t_{\gamma} \simeq 40$ms, respectively (\ref{ndt_th}b). 
In particular, we notice that the probability $P(\Delta t_{\theta}; s_c)$ is nearly independent of $s_c$ (Fig. \ref{ndt_th}b), which means  that the ratio $N(\Delta t_{\theta}; s_c)/N(s_c) \simeq const$,  and thus the number $N_{\theta}(s_c)$ of avalanches related to $\theta$ oscillations decreases proportionally to $N(s_c)$ for increasing values of $s_c$. On the other hand,  the probability $P(\Delta t; s_c)$ increases with $s_c$ for  $\Delta t > \Delta  t_{\theta}$  and decreases for $\Delta t < \Delta  t_{\theta}$, implying that 
the number of avalanches separated by longer (shorter) $\Delta t$, decreases slower (faster) than $N(s_c)$. Hence, long inter-event times  tend to separate large avalanches, whereas shorter  inter-event times tend to separate smaller avalanches.

The same analysis has been applied to a sequence in which avalanche sizes are reshuffled by keeping their occurrence time fixed. In this case
the peaks observed in  Fig. \ref{ndt_th}b disappear (inset of Fig. \ref{ndt_th}b). Since reshuffling sizes does not change  the inter-event time distribution but destroys the underlying relationship between avalanche sizes and inter-event times, we must conclude that the peaks emerging in the distributions are a consequence of the correlations between sizes and $\Delta t$s. While short quiet times and fast $\beta/\gamma$ oscillations tend to be associated with smaller avalanches, slower oscillations are in general related to larger avalanches. Indeed,  varying the threshold $s_c$ in a range of values  within  the power law regime of the size distribution $P(s)$, typically  between 30 and 400 $\mu$V (Fig. \ref{ndt_th}a), the probability $P(\Delta t_{\theta}; s_c)$ of $\Delta t$ associated with $\theta$ or slower oscillations (Fig. \ref{ndt_th}b), remains nearly unchanged. In  particular,  the coexistence of a $\theta$ peak with a faster decrease of the probability of $\Delta t_{\gamma}$ suggests an hierarchical structure in the  avalanche sequence, reminiscent of the temporal organization of nested $\theta -\beta /\gamma$ oscillations \cite{plenz:tbg} (Fig.\ref{ndt_th}c): Large avalanches occur with $\theta$ frequency and trigger smaller ones in faster $\beta/\gamma$ cycles (Fig.\ref{ndt_th}c). 
In a previous study \cite{poil12} it has been numerically shown that critical-state dynamics of avalanches and oscillations jointly emerge in a neuronal network model when excitation and inhibition is balanced. The present analysis of experimental data enlightens that oscillations in neuronal activity are the outcome of the temporal organization of avalanches with different size and provides a first empirical evidence for the coexistence of avalanches and oscillations in the critical regime of neuronal activity.

On the basis of these observations, it is clear that no universal function as in Eq.\ref{eqn:corral} can be obtained for the cortex slice data. Indeed, one can see that only the second regime of the distribution, the bell-shaped bump originated by the down-states, collapses onto a unique function when  inter-event times are appropriately rescaled  \cite{front14,epj14}.  Concerning the up-states,  the appearance of peaks indicates  that avalanche occurrence is not controlled by a unique  time scale,  and that the temporal structure is more complex.

\section{Detrended Fluctuation Analysis}

In the previous Section we have seen that a predominant characteristic of the intrinsic activity in cortex slice cultures is the continuous alternation of two distinct network states, one with prominent correlations in neural firing, the up-state, and another one with sparse and weakly correlated activity, the down-state. 
The analysis of inter-event time distributions shows that consecutive avalanches are correlated over a time-scale of about 1s, and raises the question whether avalanches separated by a longer temporal distance, namely by a down-state, are significantly correlated. In other words, we aim to understand more closely the role of down-states in the context of avalanche dynamics: Are they a sort of memory resetting period? Do they keep memory of past activity and thus correlate consecutive up-states \cite{scirep16}?

In order to address this questions, we first consider the spontaneous activity signal recorded in cortex slice cultures, $V(t) = \sum_{i=1} ^{n_e} \delta v_i (t)$ \cite{scirep16}, i.e. the sum of all potential variations recorded at time $t$ on the $n_e$ electrodes placed on the cortical slice, and apply the detrended fluctuation analysis (DFA) to quantify its long-range power-law correlations. 
The DFA \cite{peng} involves the following steps: (i) Calculate the integrated signal $I(t)=\sum_{t'=1}^{N_{max}} (V(t')-<V>)$, where $<V>$ is the mean activity and $N_{max}$ is the length of the signal; (ii) Divide $I(t)$ into boxes of equal length $n$ and fit $I(t)$ with a polynomial $I_n(t)$ of order 1, which represents the trend in that box; (iii)  Detrend $I(t)$ by subtracting the local trend, $I_{n}(t)$, in each box and calculate the root-mean-square (r.m.s.)  fluctuation $F(n)= \sqrt{1/N_{max}\sum_{t=1}^{N_{max}} [I(t)-I_{n}(t)]^2}$; (iv) Repeat this calculation over a broad range of box sizes and obtain a functional relation between $F(n)$ and $n$. 
For a power-law correlated time series, the  average r.m.s. fluctuation function $F(n)$ and the box size $n$ are connected  by a power-law relation, that is $F(n) \sim n^H$. The exponent $H$ is a parameter which quantifies the long-range power-law correlation properties of the signal. Values of $H < 0.5$ indicate the presence of anti-correlations in the time series, $H = 0.5$ absence of correlations (white noise), and $H > 0.5$ indicates the presence of positive correlations in the time series.

\begin{figure}
\begin{center}
\includegraphics[width=11.0cm]{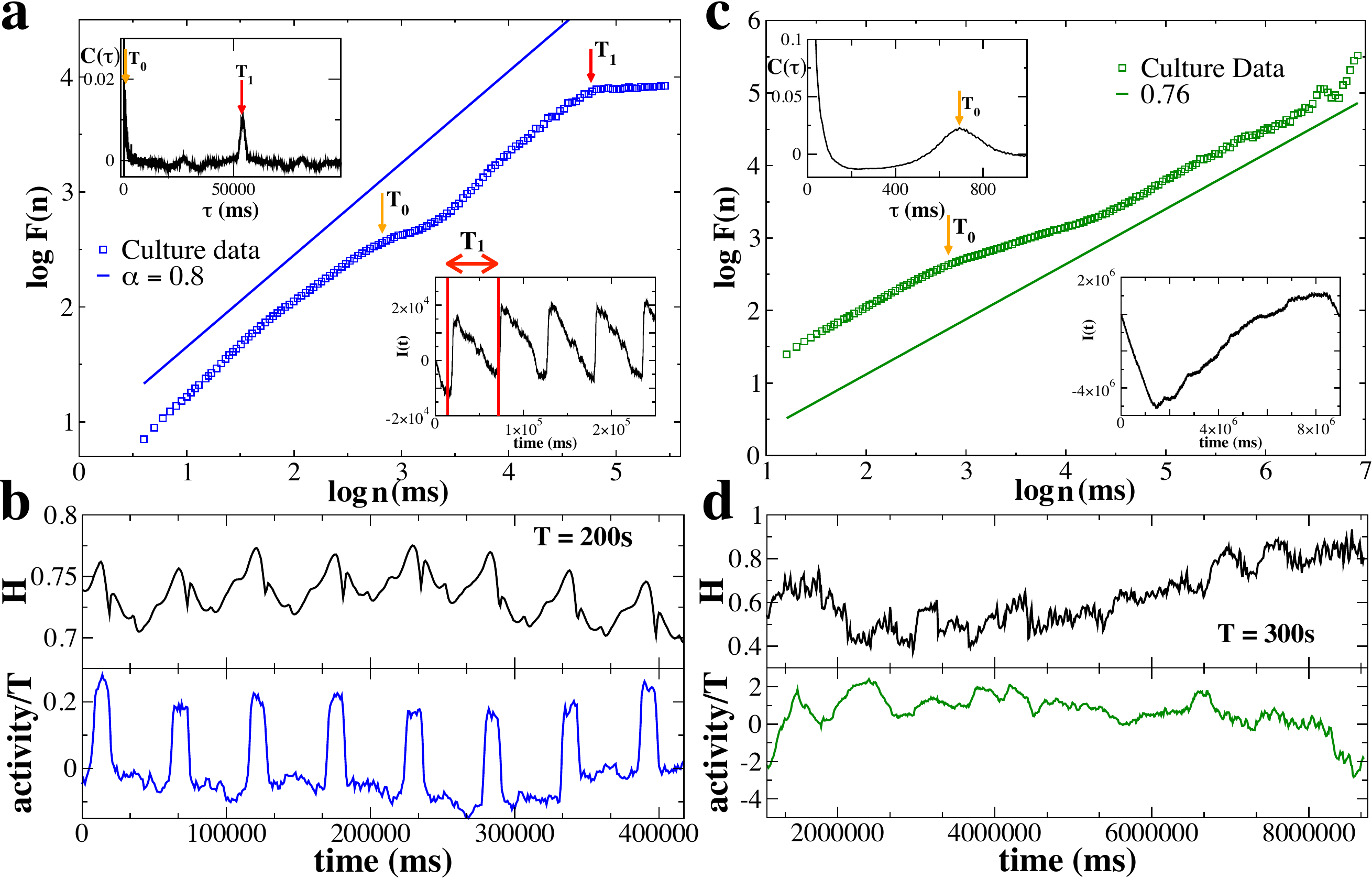}
\end{center}
\caption{Detrended fluctuation analysis of spontaneous activity for two representative samples of cortex slice cultures. The logarithm  $log F(n)$ of the r.m.s. fluctuation function is plotted versus the logarithm  $log n$ of the window size $n$. (a) $F(n)$ for the network activity $V(t)$ corresponding to the  experimental inter-avalanche time distribution (blue curve) shown in Fig. \ref{fig:ndt}. $T_0$ indicates the characteristic duration of down-states, while $T_1$ refers to a longer periodicity that may appear in the activity of cultures \cite{front14}. Upper inset: Auto-correlation of the network activity $V(t)$; Lower inset: Integrated activity $I(t)$. Vertical red lines delimit  the period $T_1$. (b) Upper panel: DFA performed using a sliding window $T = 200s$ with a step of 1s. $H$ is estimated in each window with a least square fit. Lower panel: Average network activity in consecutive sliding windows $T$. (c) $F(n)$ for the activity of a culture that only exhibits the periodicity related to the alternation of up and down-states \cite{prl12}. $T_0$ indicates the characteristic duration of down-states. Upper inset: Auto-correlation of the network activity $V(t)$; Lower inset: (d) Upper panel: DFA performed using a sliding window $T = 300s$ with a step of 1s. $H$ is estimated in each window with a least square fit. Lower panel: Average network activity in the sliding window $T$.}    
\label{dfa}
\end{figure}

For the spontaneous activity in cortex slice cultures, the fluctuation function tends to follow either of the representative behaviors that we present in Fig. \ref{dfa}.  
The first one (Fig. \ref{dfa}a) corresponds to the culture whose inter-event time distribution is discussed in Section \ref{sec:3}, and  exhibits a peculiar structure characterized by a power-law with an exponent $H \simeq 0.8$ for $n < T_0$, a crossover to a short flat region at $n \simeq T_0$, and again a power-law regime with $H \simeq 0.8$ up to $n = T_1$. For $n > T_1$, the fluctuation function  becomes completely flat.
We begin the analysis of this complex behavior by discussing the plateau for $n > T_1$. According to previous studies on the effects of specific trends on the scaling behavior of the r.m.s. fluctuation function \cite{dfa_per,dfa_nonlin}, this plateau is due to the peculiar periodic structure of the signal that can be clearly recognized in the integrated signal $I(t)$ (Fig. \ref{dfa}a, lower inset) and in the average network activity as a function of time (Fig. \ref{dfa}b, lower panel). Furthermore, the auto-correlation (Fig. \ref{dfa}a, upper inset) shows a clear peak around $T_1 = 50$s, that approximately corresponds to the length of the period observed in $I(t)$ (Fig. \ref{dfa}a, lower inset).

Inside this fundamental cycle one can easily identify a further periodicity at $T_0 \simeq 1s$ (Fig. \ref{dfa}a, upper inset), that corresponds to the characteristic duration of down-states (Fig. \ref{fig:ndt}) \cite{prl12,front14}. Following refs. \cite{dfa_per,dfa_nonlin}, this periodic trend could be responsible for  the crossover observed in the fluctuation function at $n \simeq T_0$. Indeed, this kind of crossover between two regimes with approximately the same exponent $H$, has been associated with either periodic trends \cite{dfa_per} or non-stationarities \cite{dfa_nonlin} in an otherwise long-range correlated signal, such as long segments with zero amplitude. In the signal we consider here, zero amplitude segments correspond to down-state durations and have indeed a characteristic length of about $T_0$. As pointed out in refs. \cite{dfa_per,dfa_nonlin}, the correlation features of the 'clean' signal dominate for $n << T_0$ and $n >> T_0$.     
Hence, this behavior confirms the presence of strong correlations in the up-states, i.e. $n < T_0$, and, at the same time, reveals that temporal correlations extend over a much longer range, spanning the entire period $T_1$. Interestingly, the investigation of the local behavior of the signal and its evolution over time (Fig. \ref{dfa}b), shows that the exponent $H$ fluctuates between 0.7 and 0.8 and closely follows the periodic behavior of the network activity, being larger when the network activity is more intense (Fig. \ref{dfa}b).

In Fig. \ref{dfa}c, we show a second representative behavior observed in the r.m.s.  fluctuating function of culture network activity. In this case there is only one peak in the auto-correlation at $\tau = T_0 \simeq 600$ms (Fig. \ref{dfa}c, upper inset), which corresponds to the characteristic down-state duration of the sample, and $F(n)$ exhibits a first crossover at $n \simeq T_0$ from a region with $H \simeq 0.75$ to one with $H \simeq 0.4$, followed by a second crossover to the power law regime  with $H \simeq 0.75$ (Fig. \ref{dfa}c). As for the first representative case (Fig. \ref{dfa}a), we observe that the short and long timescale  behavior of the fluctuation function are very similar, which suggests that the network activity features long-range  temporal correlations transcending up-state durations. 
Similarly to the scenario in Fig. \ref{dfa}a, here the short and long-range power law scaling of $F(n)$ are separated by a crossover that starts around $n \simeq T_0$. However, the interpretation of  this extended crossover to a region with $H \simeq 0.4$ is more delicate. 
On the one hand, it  seems to be related to the down-state duration $T_0$, as discussed above. On the other hand, the analysis of the local behavior of the signal (Fig. \ref{dfa}d), evidences the presence of anti-correlated segments with  $H \simeq 0.4$ (Fig. \ref{dfa}d, upper panel). To better understand the origin of the crossover at $n \simeq T_0$, we have repeated the DFA in segments of the signal with and without anti-correlations. In both cases, we have found a very similar behavior for $10^3<n<10^4$ as in the $F(n)$ averaged over the entire signal, which suggests that again the crossover at $n \simeq T_0$ has to be associated with the rhythmic alternation  between up and down-states.

In summary, although the presence of periodic trends and non-stationarities  gives rise to crossovers and non-constant scaling exponents, the DFA   indicates that temporal correlations extend beyond the duration of a single up-state and transcend the down-state time scale. As suggested by numerical simulations, the crossover located around the characteristic time of the down-state may be indicative of a change in the nature and intensity of correlations in the down-state. To better understand this last point and provide a  more refined characterization of the temporal structure of the intrinsic activity, in the following Section we introduce a method to systematically investigate the relation between avalanches and inter-times.

\section{Conditional probability analysis}

The DFA represents a preliminary approach in the query for temporal correlations in neuronal signals. As we have seen above, DFA can provide information about the nature of the correlations and help in identifying distinct dynamical mechanisms  that operate on specific timescales. However, this approach does not allow to investigate the temporal organization  of the process, namely the correlations between the avalanche sizes and their occurrence times. In order to address this question, we apply a powerful statistical method to the temporal sequence of neuronal avalanches, where avalanches are characterized by their size $s_i$, their  starting and ending time, $t_i^i$ and $t_i^f$, respectively. 

The method, developed for the analysis of seismic catalogs \cite{prl100}, is based on the analysis of conditional probabilities on real and surrogate data. 
Specifically, here we evaluate the probability $P(s_{i+1}/s_i>\lambda|\Delta t<t_0)$  to observe two consecutive avalanches with size ratio $s_{i+1}/s_i>\lambda$ under the condition that their temporal separation  $\Delta t=t_{i+1}^i-t_i^f<t_0$, where $\lambda$ and $t_0$ are parameters. For each couple of parameter values, we evaluate the same probability in temporal sequences where avalanche sizes are reshuffled in time by keeping their occurrence times fixed. Since in reshuffled sequences avalanche sizes are uncorrelated by construction,  the conditional probability $P$ evaluated for several realizations of reshuffled sequences follows a Gaussian distribution, whose average value $P^*$ and standard deviation $\sigma$ depend on the parameters $\lambda$ and $t_0$ \cite{scirep16}.
Then, we compare  $P(s_{i+1}/s_i>\lambda|\Delta t<t_0)$ with $P^*$ by considering the difference $\delta P=P(s_{i+1}/s_i>\lambda|\Delta t<t_0)-P^*$ for each couple of parameters. If $|\delta P > 2\sigma|$, we conclude that  non-zero correlations exist between the size of avalanches separated by  $\Delta t<t_0$, and distinguish two cases: $\delta P > 0$ and $\delta P < 0$. In the first case, it is more likely to observe two consecutive avalanches distant in time $\Delta t<t_0$ with a size ratio $s_{i+1}/s_i>\lambda$ in the original avalanche sequence rather than in a reshuffled sequence. This indicates that consecutive avalanche sizes are positively correlated. Conversely, we say that consecutive avalanches are anti-correlated if $\delta P < 0$, meaning that it is more likely to observe two consecutive avalanches distant in time $\Delta t<t_0$ with a size ratio $s_{i+1}/s_i>\lambda$ in a  random sequence rather than in the original one.

\begin{figure}
\begin{center}
\includegraphics[width=11.5cm]{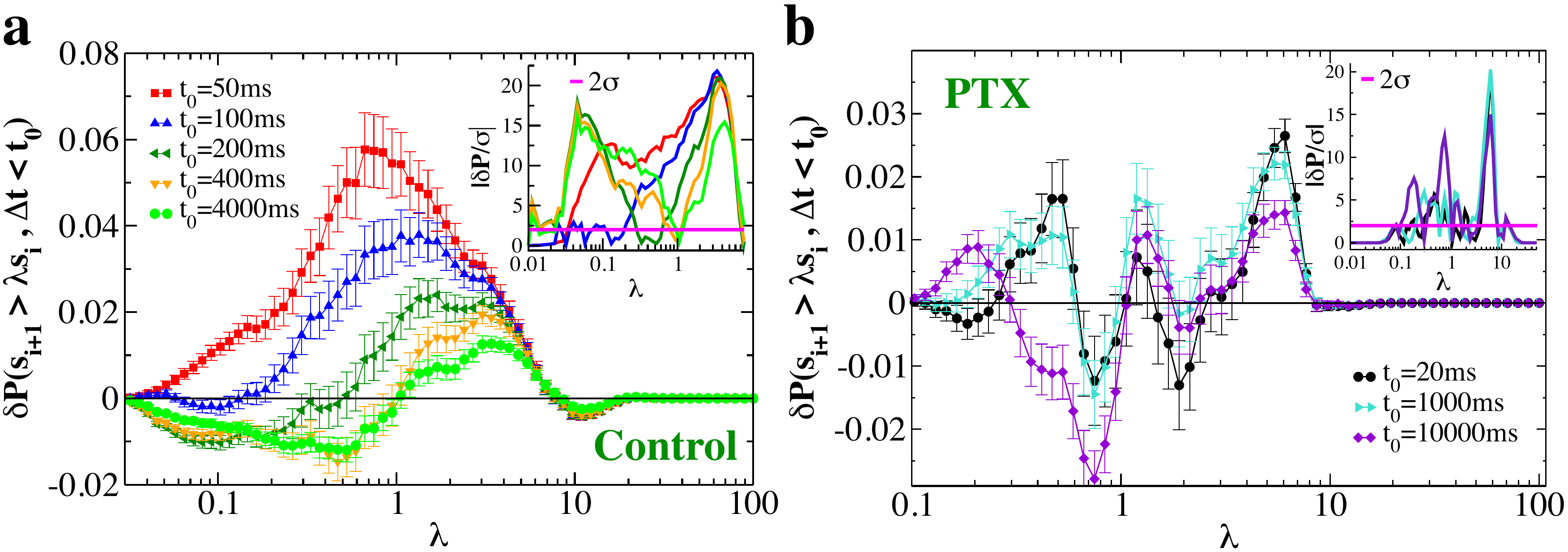}
\end{center}
\caption{(a) The quantity $\delta P(s_{i+1} > \lambda s_i, \Delta t_i < t_0)$ as a function of $\lambda$ for different values of $t_0$ in normal conditions. The bar on each data point is $2\sigma$. Each curve represents an average over all experimental samples. Inset: The ratio $|\delta P|/\sigma$ as a function of $\lambda$ for different values of $t_0$. The magenta horizontal line indicates the value $2\sigma$. (b) The same quantities for disinhibited cultures treated with PTX.}   
\label{fig:deltaP}
\end{figure}
 
In Fig.\ref{fig:deltaP}a we plot $\delta P$ averaged over seven samples of cortex slices as function of $\lambda$ for different values of the temporal distance $t_0$. Data show that the intensity of correlations depends on $t_0$ since curves clearly separate and the amplitude decreases if farther in time avalanches are included in the probability evaluation. The maximum of the conditional probability indicates the most probable size ratio for different $t_0$. Interestingly the same analysis performed for other stochastic processes, as earthquakes and solar flares, provides a maximum position not changing as $t_0$ increases. Conversely, we observe that, if two consecutive avalanches have a temporal distance smaller than 100ms, the maximum is located at $\lambda<1$, indicating that the second avalanche tends to be smaller than the first one. By increasing the temporal distance, the maximum drifts towards the region $\lambda>1$, implying that  for consecutive far-in-time avalanches, the second one tends to be larger. 

This complex temporal organization can be understood in the context of  the alternation of up and down-states pointed out in Section \ref{sec:3}. In the up-state avalanches are close in time and their sizes are correlated. Due to the longstanding activity, avalanche sizes tend to gradually decrease. Conversely, far-in-time avalanches occur after a down-state, during which the system is able to recover resources and then trigger larger avalanches.

Crucially, for disinhibited cultures (PTX) this temporal organization is strongly altered \cite{scirep16}, and individual cultures may exhibit either single or multiple peaks in $\delta P$. As a consequence,  the group average shown in Fig. \ref{fig:deltaP}b exhibits  multiple peaks located at different values of $\lambda$.  Importantly, all analyzed cultures exhibit a relative maximum for $\lambda > 1$ that is nearly independent of $t_0$, evidencing that in the disinhibited condition, the avalanche process is extremely unbalanced, and an avalanche can be considerably larger than the previous one, independently of their  time separation. These observations further suggest that the temporal structure of neuronal avalanches is the signature of a healthy system, and is controlled by fundamental biomolecular mechanisms.

\begin{figure}
\begin{center}
\includegraphics[width=11.0cm]{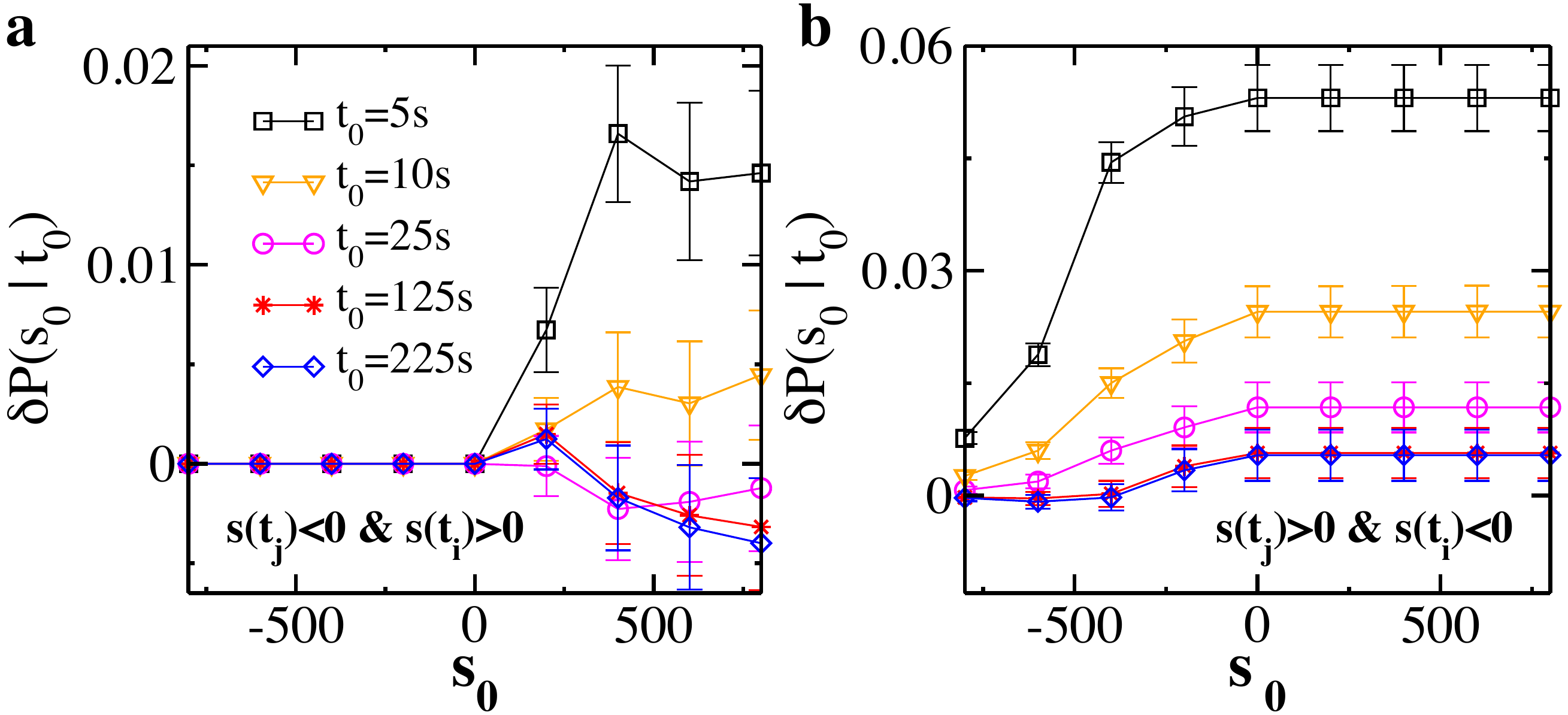}
\caption{The quantity $\delta P(s_0, t_0)$ as a function of $s_0$ for $t_0 = 5, 25, 125, 175, 225$ s. For each $t_0$ and $s_0$ the error bar is the standard deviation $\sigma(s_0, t_0)$. Different panels consider different combinations of successive variations with opposite sign.}
\end{center}   
\label{fig:csf}
\end{figure}

The same analysis can be performed on  neuronal signals from systems at larger scale, for instance to study correlations between large events in the brain blood oxygenated level dependent (BOLD) signal measured by fMRI on healthy patients \cite{csf}. The activity $B(\vec r_i, t)$ is monitored at each voxel $i$ as function of time. Data are recorded in time every $\delta t = 2.5$ s, therefore time is measured in units of $\delta t$. The study focuses only on extreme activity events and therefore analyzes voxels for which  $B(\vec r_i, t)$  is larger than a given threshold $B_c = 18000$, value that selects the largest 10\% of the entire activity range. The below-threshold values are substituted by zero. Rather than the activity itself, the interesting quantity is the activity variation at each voxel, $s_i (t)= B(\vec r_i, t+ \delta t)-B(\vec r_i, t)$, which can have positive and negative values. Following the same procedure outlined before, $\delta P=P(\Delta s<s_0|\Delta t<t_0)-P^*$ measures the probability to observe the difference between successive variations occurring  at any couple of voxels $l$ and $m$, $\Delta s= s_m(t')-s_l(t)<s_0$, at a temporal distance $\Delta t=t'-t<t_0$, compared to a reshuffled, uncorrelated sequence. To provide a more detailed information, data can be analysed by separating the four different cases of successive variations having the same/opposite sign. This study evidences  \cite{csf} that successive variations with the same sign (both positive or both negative) are anti-correlated over a short time scale, i.e. less than few seconds, and they are uncorrelated over longer time scales.  Conversely,  consecutive variations with opposite sign appear to be strongly correlated over the time scale of few seconds (Fig. \ref{fig:csf}). The overall analysis suggests that successive activity enhancements or depressions are very unlikely if close in time, and completely uncorrelated over longer temporal scales. Conversely, decreasing activity in some voxels triggers activity enhancements in other voxels after a short time delay, and vice versa. Variations of different signs show, indeed, a strong evidence of correlations suggesting that the system activity realizes a sort of homeostatic balance compensating local enhancements and depressions.

\section{Conclusions}

In this review we have presented a number of statistical analyses of experimental data aimed  at understanding avalanche dynamics in neural networks through the study of temporal correlations. We focused on spontaneous local field potential activity recorded in cortex slice cultures, and on the resting fMRI  BOLD signal. 
In order to achieve a deeper comprehension of the microscopic mechanisms leading to the temporal organization observed at the macroscopic scale, we compare experimental results to numerical simulations of a network of integrate and fire neurons. 
A first achievement of numerical simulation was the identification of the crucial role of inhibition in the scaling behavior of the EEG power spectra. 
We have shown that the PSD follows a power-law whose exponent scales with the percentage of inhibitory synapses, and tends to $1/f$ noise for a percentage of about 30\%, a value usually estimated for mammal brains. Conversely, fully excitatory networks exhibit $1/f^2$ PSD, i.e. brown noise, which suggests that long range correlations can be obtained only if a certain level of inhibition is present in the network \cite{chaos}. These evidences indicated that the frequency spectrum of resting brain activity may be controlled by the ratio between excitation and inhibition, and could be used as a tool to measure this ratio experimentally, as more recently shown in  \cite{ei_balance}, and  thus investigate pathological conditions. Indeed the analysis of the PSD in epileptic patients provide a scaling exponent close to 2, while in healthy subjects it generally falls in the interval [0.8,1.4].

We further investigated the role of balanced excitation and inhibition  focusing on the sequence of avalanches \cite{prl12,front14} in normal and disinhibited condition. Within this approach, avalanche activity is considered as a point process and attention is shifted to the time elapsed between successive avalanches, i.e. the inter-event time. The analysis of  inter-event times is widely used to investigate temporal correlations, which are often associated with the presence of a power-law regime in the inter-event time distribution. For instance, in seismology this distribution behaves as a Gamma function, where the initial power law decay is the expression of correlations in earthquake occurrence due to aftershock sequences triggered by large mainshocks. 
In spontaneous activity of rat cortex slice cultures, we consistently observe a power-law regime that indicates significant temporal  correlations between consecutive avalanches over a temporal range of several hundreds of milliseconds. However, the general functional  behavior is more complex than a Gamma function,  since the distribution exhibits non-monotonicity and a characteristic inter-event time of about one second. 
Numerical simulations are able to provide  insights into this peculiar behavior, and show that the underlying  avalanche dynamics results from the alternation of up and down-states, corresponding to long bursty and quiescent periods, respectively. The crucial observation is that data from different samples are reproduced by numerical simulations if the ratio of the two parameters controlling those two distinct network states is tuned to the same, "optimal" value. This ratio controls the dynamic balance of excitation and inhibition that is realized  by the alternation of high and low activity periods, and goes beyond the  structural inhibition, i.e. the percentage of inhibitory synapses. The combined influence of dynamical and structural inhibition on the PSD of the network activity will be subject of upcoming studies.  
In this review we have shown that unbalanced excitation disrupt the temporal organizazion of avalanches and alters the functional form of the inter-avalanche  time distribution, at the same time moving the system  away from criticality. 
These conclusions are supported by a remarkable agreement with ours and other  experimental results \cite{shew09}. Indeed, the comparison of critical and disinhibited, supercritical cultures evidences the same changes observed in numerical simulations \cite{scarpetta2014}. 

The analysis of inter-event time distributions has also shown that avalanche occurrence preserves the temporal features of $\theta$ and $\beta/\gamma$ oscillations \cite{front14}, and has uncovered a hierarchical structure where large avalanches  occurring  with $\theta$ frequency trigger cascades of smaller avalanches corresponding to the higher frequency oscillations, reminiscent of the temporal organization of nested $\theta -\beta /\gamma$ oscillations \cite{he2010}. Remarkably, our analysis showed that characteristic brain rhythm time scales do not imply characteristic avalanche sizes, and it indicates that different rhythms interact and organize in time as avalanches with scale-free size.
The connection between nested oscillations and neuronal avalanches has been also  pointed out in \cite{plenz:tbg}. Investigation of spontaneous neuronal activity in the rat cortex layer 2/3  has shown that  bursts develop  a temporal organization of higher frequency oscillations, $\beta$ and  $\gamma$, nested into  $\theta$  oscillations, while the spatio-temporal organization of LFPs is characterized by the  scaling behavior of neuronal avalanches.

The self-regulated balance of excitation and inhibition is widely considered as a fundamental property of 'healthy' neural systems. Pharmacological  alterations of this balance can drive  those systems to a pathological conditions and thus away from criticality, as seen above. Such a  kind of perturbation not only influences the avalanche statistics, as previously reported \cite{shew09}, but determines a general reorganization of avalanches in time, as we discussed here. In order to address this point, we used a technique that compares conditional probabilities evaluated in real and reshuffled time series, and it is particularly suitable for understanding the relationship between bursty and quiescent periods \cite{physrep,scirep16}. 
In a recent study we have shown that suppressing inhibition disrupts the relationship between bursts and quiescence that characterizes cortical networks at criticality. In particular, while in the critical state longer recovery periods are needed for the system to generate large avalanches, in the disinhibited condition large avalanches also arise after short periods of quiescence \cite{scirep16}. As a consequence, in critical cultures an avalanche tends to be smaller  than the preceding one in the up-states, i.e. for $\Delta t$ shorter than about 200 ms, and viceversa when the time separation is larger, namely when two avalanches are separated by a down-states. In the case of network disinhibition, this organization is strongly altered and an avalanche can be considerably larger than the previous one, independently of their time separation.

Overall, the study of inter-event time and their relation with avalanche sizes reveals basic features of avalanche dynamics: Temporal correlations are certainly relevant in the up-states, and the analysis of conditional probabilities indicates  that also avalanches belonging to consecutive up-states, namely separated by a down-state,  are correlated, although the nature and strength of their correlation may be different. The detrended fluctuations analysis  confirms these conclusions and indicates that network activity is long-range correlated. The r.m.s. fluctuation function is generally characterized by the same scaling exponent at short and long timescale, $H \simeq 0.8$, and these two power law regimes are separated by a crossover to a  region with $H \simeq 0.5$. This crossover is always located around the characteristic time of the down-state and may  be a consequence of a change in the nature and intensity of correlations in this network state \cite{dfa_per,dfa_nonlin}.

Finally, we have shown that balanced network activity can be also recognized at larger scale, in the fMRI BOLD signal \cite{csf}.  Variations in the signal at each voxel exhibit clear correlations in time: A local increase in activity is followed by a close in time decrease in activity at other voxels and vice-versa. Therefore, at different scales, for different neuronal systems and signal kinds, results consistently indicate that  healthy neuronal systems  self-regulate the dynamic balance of excitation and inhibition, which is the expression of relevant temporal correlations in network activity. 
Balanced networks exhibiting critical avalanche dynamics are able to  accomplish tasks, as learning Boolean rules and multi-task learning, characterized by properties observed in real systems \cite{pnas,opt}. Criticality could therefore play an important role in the optimal  response to external stimuli, in information processing, and in the functional performances of neuronal systems in general.

\end{document}